\documentclass[aps,prx,twocolumn,superscriptaddress,a4paper,floatfix,amsfonts,amsmath,amssymb,citeautoscript]{revtex4-2}

\usepackage{mathptmx}
\usepackage{float}
\usepackage[scaled=0.9]{helvet}
\usepackage[utf8]{inputenc}
\usepackage{graphicx}
\usepackage[pdftex,dvipsnames]{xcolor}
\usepackage[pdftex]{hyperref}
\hypersetup{
    pdftitle={},
    colorlinks,
    citecolor=blue
    }
\usepackage{upgreek}
\usepackage{bm}
\usepackage{textcomp}
\usepackage[
,textwidth=17.5cm
,textheight=23.5cm
,verbose
,dvips
]{geometry}
\usepackage{xspace}
\usepackage{xargs}
\usepackage{verbatim}

\usepackage[normalem]{ulem}
\usepackage{ifthen}

\definecolor{gray}{rgb}{.6,.6,.6}
\definecolor{darkyellow}{rgb}{.6,.5,0}
\definecolor{darkgreen}{rgb}{0,.6,0}
\definecolor{darkred}{rgb}{.6,0,0}

\DeclareRobustCommand{\void}[1]{}

\newcommand{\fig}[2]{Fig.~\ref{#1}\ifthenelse{\equal{#2}{}}{}{(\lowercase{#2})}}
\newcommand{\Fig}[2]{Figure~\ref{#1}\ifthenelse{\equal{#2}{}}{}{(\lowercase{#2})}}

\newcommand{\eq}[1]{Eq.~\eqref{#1}}

\newcommand{\sect}[1]{Sec.~\ref{#1}}
\renewcommand{\paragraph}[1]{{\par\it #1.---}\ignorespaces}

\newcommand{\app}[1]{Appendix \ref{#1}}

\newcommand{\fsaw}{\ensuremath{f_\text{SAW}}}
\newcommand{\lsaw}{\ensuremath{\lambda}}

\newcommand{\Omegar}{\ensuremath{\Omega_\text{r}}}
\newcommand{\Omegai}{\ensuremath{\Omega_\text{i}}}

\begin{document}

%\title{In-operando spatially resolved strain-dynamics in focusing microcavities as studied by scanning X-ray diffraction microscopy}
\title{Scanning X-ray diffraction microscopy of a  6\,GHz surface acoustic wave}

\author{M. Hanke}
\email{hanke@pdi-berlin.de; ludwig@pdi-berlin.de}
\affiliation{Paul-Drude-Institut für Festkörperelektronik, Leibniz-Institut im Forschungsverbund Berlin e.V., Hausvogteiplatz 5--7, 10117 Berlin, Germany}
\author{N. Ashurbekov}
\affiliation{Paul-Drude-Institut für Festkörperelektronik, Leibniz-Institut im Forschungsverbund Berlin e.V., Hausvogteiplatz 5--7, 10117 Berlin, Germany}
\author{E. Zatterin}
\affiliation{ESRF, European Synchrotron, 71 Ave Martyrs, F-38000 Grenoble, France}
\author{M.\,E. Msall}
\affiliation{Department of Physics and Astronomy, Bowdoin College, Brunswick, Maine 04011, USA}
\author{J. Hellemann}
\affiliation{Paul-Drude-Institut für Festkörperelektronik, Leibniz-Institut im Forschungsverbund Berlin e.V., Hausvogteiplatz 5--7, 10117 Berlin, Germany}
\author{P.\,V. Santos}
\affiliation{Paul-Drude-Institut für Festkörperelektronik, Leibniz-Institut im Forschungsverbund Berlin e.V., Hausvogteiplatz 5--7, 10117 Berlin, Germany}
\author{T.\,U. Schulli}
\affiliation{ESRF, European Synchrotron, 71 Ave Martyrs, F-38000 Grenoble, France}
\author{S. Ludwig}
\email{hanke@pdi-berlin.de; ludwig@pdi-berlin.de}
\affiliation{Paul-Drude-Institut für Festkörperelektronik, Leibniz-Institut im Forschungsverbund Berlin e.V., Hausvogteiplatz 5--7, 10117 Berlin, Germany}

\begin{abstract}
Surface acoustic waves at frequencies beyond a few GHz are promising components for quantum technology applications. Applying scanning X-ray diffraction microcopy we directly map the locally resolved components of the three-dimensional strain field generated by a standing surface acoustic wave on GaAs with wavelength $\lambda\simeq500\,$nm corresponding to frequencies near 6\,GHz. We find that the lattice distortions perpendicular to the surface are phase-shifted compared to those in propagation direction. Model calculations based on Rayleigh waves confirm our measurements. Our results represent a break through in providing a full characterization of a radio frequency surface acoustic wave beyond plain imaging.
\end{abstract}

\maketitle

\section{Introduction}\label{sec:introduction}

Synchrotron-based scanning X-ray diffraction microscopy (SXDM) is a non-destructive tool to directly measure the local strain distribution of nanoscale devices \cite{DMB11,KHN17,HLK18,LCD19} and to explore nanoscale strain dynamics operando \cite{AKR19,SHK17}. In the past, it was applied to visualize and probe the time resolved dynamics of surface acoustic waves (SAWs) at relatively long wavelengths \cite{SSM99,ZoQ02,WAB19,IrR14,IEK15,RSB13}.

SAWs, already studied in 1885 by Lord Rayleigh \cite{Rayleigh1885}, are infamous for being the most destructive seismic waves originated by earthquakes. SAWs of much smaller wavelengths are important components of integrated circuits serving as miniature radio-frequency (rf) filters \cite{Ruppel2017,Mahon2017}, sensors \cite{Kalinin2011,Devkota2017}, or for micro controlling fluids \cite{Wixforth2006,Ding2013}. While these devices already in the market are based on SAWs with wavelengths beyond a few micrometers, SAWs at sub-micrometer wavelengths are required for new applications in quantum technology \cite{Delsing2019}, for instance, for high-quality cavities of phonons with energies beyond $\sim10\,\mu$eV \cite{Buyukkose2013}. Such localized phonons could be used as coherent on-chip interconnects between quantum bits or as components of hybrid quantum bits \cite{Forster2014,Schuetz2015}.

Methods often used to characterize SAWs at lower frequencies include electrical rf reflection or transmission measurements \cite{Buyukkose2013}, laser interferometry \cite{Santos2018} or atomic force microscopy \cite{HMM22}. For high frequency SAWs with sub-micrometer wavelengths, the electrical measurements suffer from impedance matching problems and the typically small amplitudes of high frequency SAWs. While laser interferometry has an insufficient lateral resolution \cite{Santos2018} at wavelengths well below $\lambda<1\,\mu$m, atomic force microscopy in principle has the necessary lateral and vertical resolution needed for imaging rf standing SAW (SSAW) modes. However, none of the above methods can provide a full characterization of a SAW. In contrast, SXDM can not only resolve rf SSAWs but is sensitive to all components of the three-dimensional local strain field, the essence of a SAW.

At the ID01 beamline of the European Synchrotron Radiation Facility (ESRF) \cite{CRH14,CZR15,LFZ17,LCD19}, we have performed SXDM measurements mapping the local strain field of an SSAW realized on a GaAs wafer at a phonon wavelength of $\lsaw\simeq500\,$nm corresponding to a frequency near $f=6\,$GHz. Comparing our findings with the theory of Rayleigh waves \cite{Rayleigh1885,Schuetz2015} we demonstrate a remarkable agreement between the measured lattice distortions and theory predictions, which includes a characteristic phase shift between the longitudinal versus transversal SAW components.

\section{Setup and Experimental Techniques}\label{sec:setup}

\begin{figure}[th]
\includegraphics[width=1\columnwidth]{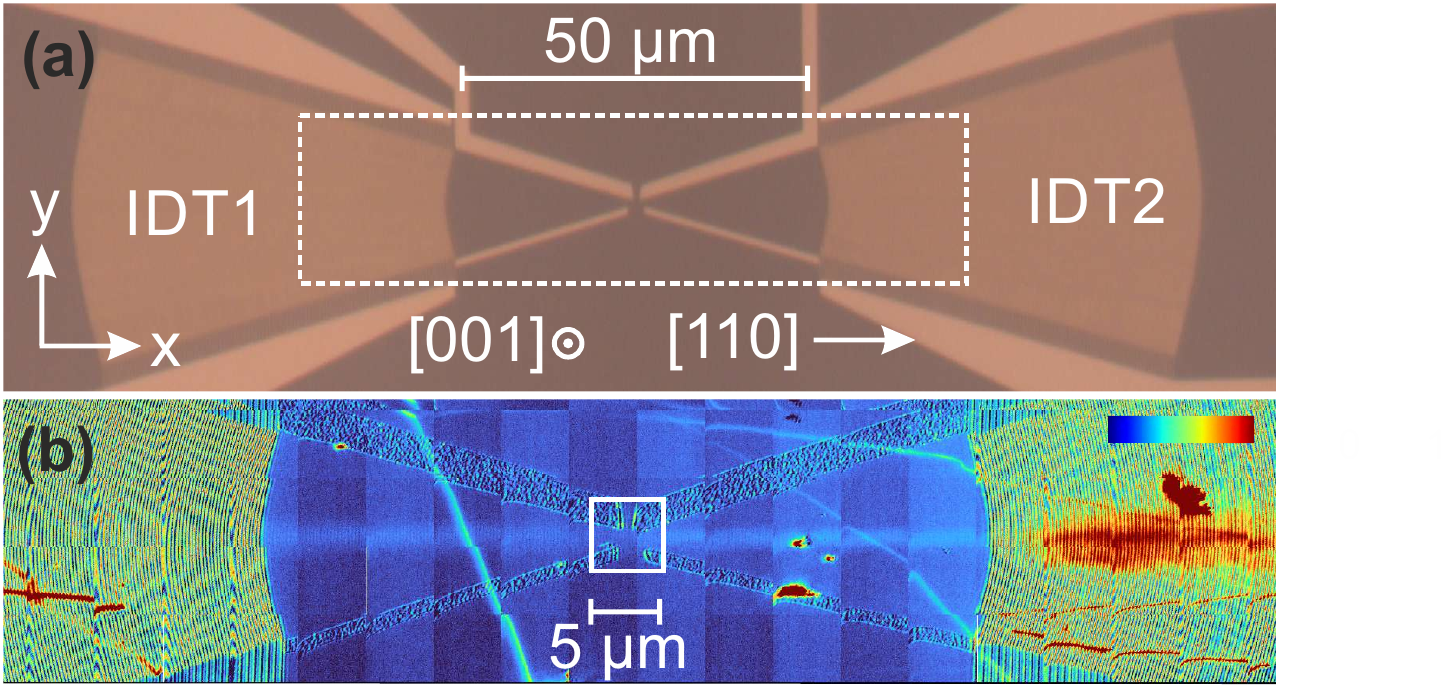}
\caption{(a) Optical micrograph of the SAW cavity consisting of two focusing IDTs with a pitch of 250\,nm (not resolved) and additional guard gates in the center. The Ti/Al/Ti (10\,/30\,/10\,nm) gates (including the IDTs), which are fabricated by electron beam lithography, are depicted by a lighter tone on the GaAs surface with a darker color. (b) The color scale depicts, within the highlighted rectangular region in (a), the diffusely scattered intensity of the symmetric GaAs(004) reflection of an SXDM measurement. Dark-blue corresponds to the intensity of the unstrained crystal, red depicts enhanced intensity due to strain. The data are assembled of successively measured individual scans within apparent $5\times5$\,$\mu$m$^2$ squares. The metal gates including the IDTs are clearly visible. An rf-excitation of IDT2 at 5.690\,GHz causes lattice distortions in the GaAs substrate underneath (red), which excite an SSAW inside the cavity along $\left<110\right>$ seen as a horizontal light-blue stripe.}
\label{fig:sample}
\end{figure}
Our SSAWs are confined in acoustic microcavities aligned along the $\langle110\rangle$ direction on a GaAs (001) surface. The cavities, designed for quantum applications based on the coherent interaction between electrons and phonons \cite{Schuetz2015}, are shaped by two mirror-symmetric focusing interdigital transducers (IDTs), cf.\ \fig{fig:sample}{a} \cite{Santos2018,Delsing2019}. To generate an SSAW, we modulate the right-hand-side IDT gates with an electrical rf-signal, while the finger gates of the left-hand-side IDT are grounded, such that it functions as a passive Bragg mirror. In \fig{fig:sample}{b} we present a typical large-scale scanning SXDM image of the cavity including an SSAW mode at $\fsaw=5.690\,$GHz. The color scale depicts the intensity caused by diffusive scattering collected near the  GaAs(004) reflection but excluding the intensity maximum. It expresses the local strain distribution perpendicular to the surface within the GaAs crystal. (Here, the strain includes local tilts of the surface.) Evident strain patterns below the evaporated metal gates allow an easy orientation on the surface. The SSAW appears as a horizontal stripe of enhanced strain (brighter color) centered between the IDTs. The region of maximal strain (red color) below the driven IDT is the source of the excited SSAW mode. Various additional speckles and stripes of high strain point to defects near the surface.
\begin{figure}[th]
\includegraphics[width=0.9\columnwidth]{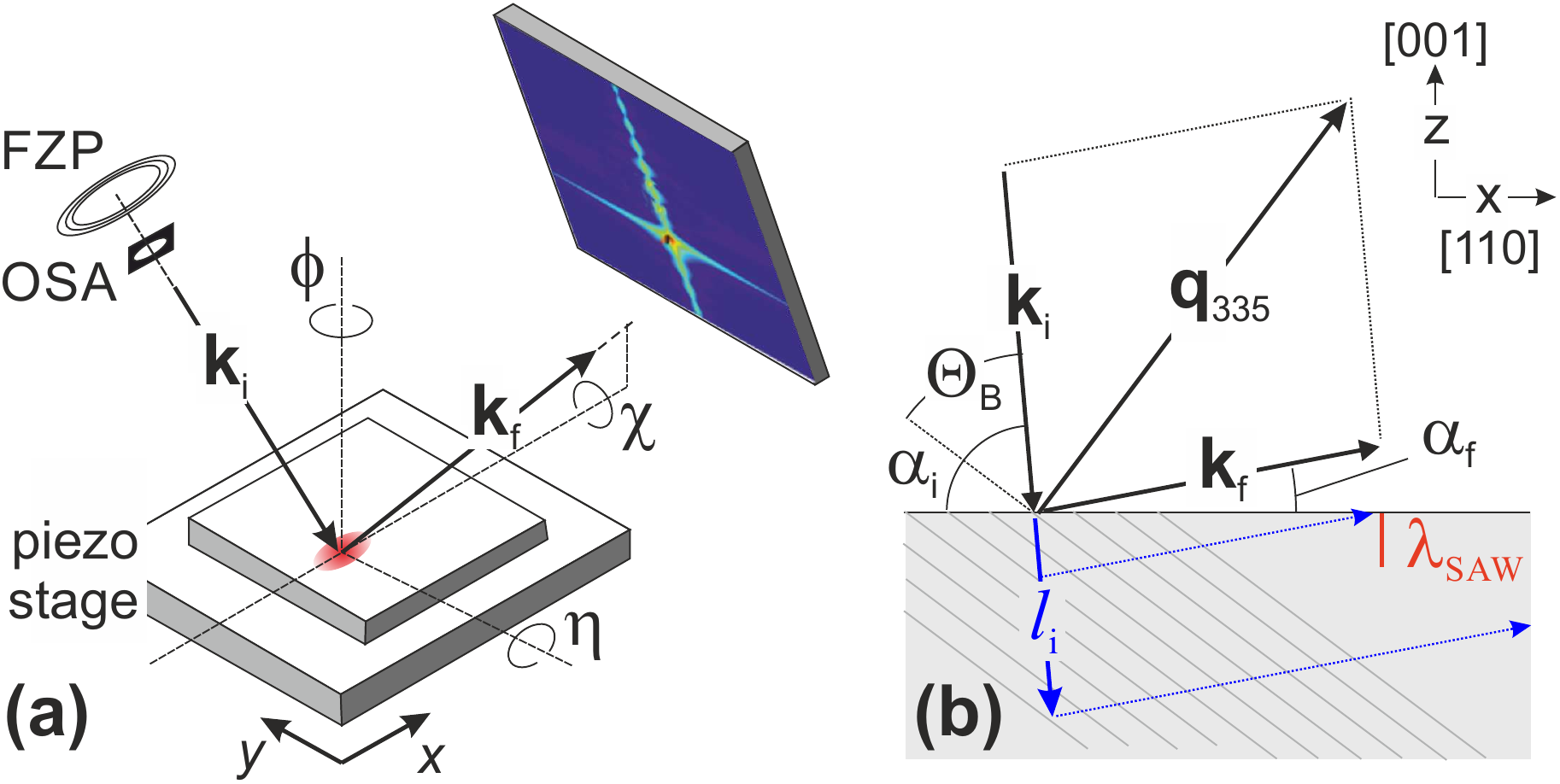}
\caption{Experimental setup (a) and coplanar scattering geometry (b). X-ray focusing is achieved by a combination of a 300\,$\mu$m gold Fresnel zone plate (FZP) with an outermost zone width of 20\,nm and an order sorting aperture (OSA). Samples are mounted on a piezo scanning stage for precise three-dimensional translations, while a hexapod on a rotation axis underneath (not shown) provides macroscopic sample translations and three orthogonal rotation axes ($\varphi$, $\chi$ and $\eta$). The scattered intensity is recorded by a 2D MAXIPIX detector. In order to maximize the spatial resolution we decided for a strongly asymmetric scattering geometry with nearly normal incidence.}
\label{fig:setup}
\end{figure}
The SXDM setup is sketched in \fig{fig:setup}{a}. It shows the $x$-$y$-$z$ piezoelectric scanning stage, which in turn is fixed to a moveable hexapod serving for controlling the orientation and rough position of the sample mounted to the piezo stage. Using a Fresnel zone plate, an X-ray beam with an energy of 10.185\,keV is focused onto the sample at a focal length of 20\,mm, a lateral spot size of $\simeq(30\,\text{nm})^2$, and a primary intensity of $\simeq5\times 10^9$ photons per second.

A Rayleigh wave is expected to cause periodic crystal displacements in its propagation direction parallel to the surface, in our setup along $\left<110\right>$ ($x$-axis) and perpendicular to the surface, along $\left<001\right>$ ($z$-axis). The penetration depth of a Rayleigh wave is comparable to its wavelength, here $\lambda\simeq500\,$nm. In order to probe the related modulations of the diffraction vector $\delta\mathbf{q}$ in these two directions, we select the asymmetric (335) reflection of GaAs as $\mathbf{q}_{335}$ is composed of similar components $q_x=4.713\,\text{\AA}^{-1}$ and $q_z=5.557\,\text{\AA}^{-1}$. The resulting diffraction geometry is co-planar, such that the scattering plane spanned by the incoming and diffracted wave vectors $\textbf{k}_\text{i,f}$ is orthogonal to the crystal surface. The incidence of the X-rays is for $\mathbf{q}_{335}$  almost vertical with $\alpha_i=85.2^\circ$, cf.\ \fig{fig:setup}{b}. This enables a high lateral resolution due to a small beam footprint. It also leads to a small exit angle of $\alpha_f=4.6^\circ$, which causes a noticeable damping of the scattered X-rays due to their path through the crystal almost parallel to the surface. This damping and the extinction depth of the incoming X-rays in GaAs of $l_\text{ext}\simeq4.6\,\mu$m, both, cause a reduction of the contribution to $\delta\mathbf{q}$ as $z$ increases.

The planar screen sketched in \fig{fig:setup}{a} shows a detector frame, which displays a reflection in reciprocal space measuring an individual spot on the surface of the fixed sample. The reflection is  distorted and broadened according to the imaging function of the focused X-ray beam, including a minor contribution due to the mapping of a spherical diffraction pattern onto a planar screen. Importantly, this distortion of the X-ray beam does not affect a relative shift of the reflection because of the $\delta\mathbf{q}$ caused by an SSAW.

To obtain the complete 3D reflection, we measure a rocking curve by varying the Bragg angle $\Theta_\text{B}$, cf.\ \fig{fig:setup}{b}, by $\pm0.2$°, which corresponds to $\eta$ in \fig{fig:setup}{a}. Scanning the sample through the $xy$-plane then yields a set of 3D detector images mapping the sample surface (in real space). In summary, a complete surface scan consists of a 5D data set. It is composed of 3D images of the time-averaged intensity of the GaAs(335) reflection, while the 2D scan in real space provides its spatial variations. Applying an SSAW results in time-periodic spatial variations $\delta\mathbf{q}$ of the scattering vector, modulating the position of the reflection. Practically, the SSAW induced $\delta\mathbf{q}$ is small compared to the width of the undisturbed broadened reflection. Moreover, our measurement averages in time over many SSAW periods and we expect $\left<\delta\mathbf{q}\right>_t\simeq0$ for moderate SSAW powers within a linear response regime. To nevertheless analyze the effect of the SSAW, we fit the scattered intensity around the GaAs(335) reciprocal lattice point (the reflection peak) with three orthogonal Gaussians and determine their individual standard deviations $\sigma_{x}$, $\sigma_{y}$, and $\sigma_{z}$.

\section{Results and Discussion}\label{sec:results}

\begin{figure*}[t]
\includegraphics[width=1.0\textwidth]{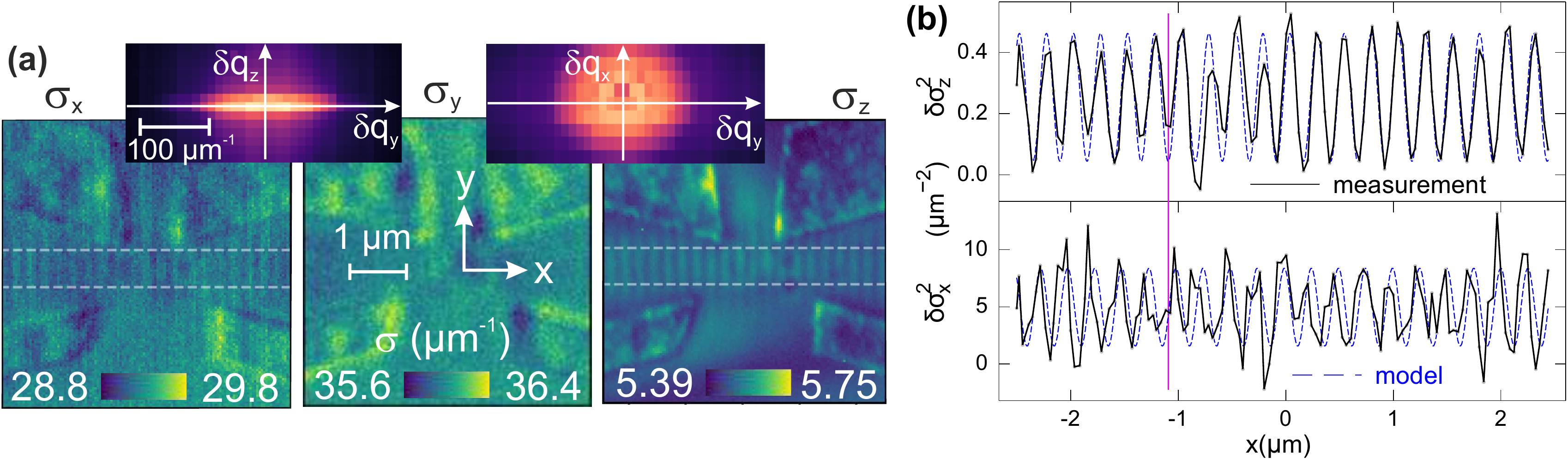}
\caption{(a) The upper plots show two orthogonal projections of an exemplary GaAs(335) reflection.  The undisturbed widths are $(\sigma_0)_x=29.1\,\mu$m$^{-1}$ and $(\sigma_0)_z=5.5\,\mu$m$^{-1}$. The color scales of the lower plots depict the standard deviations, $\sigma_x$, $\sigma_y$ and $\sigma_z$ of Gaussians fitted to the reflection along the respective directions, while we scanned the center of the cavity inside the square in Fig.\,\ref{fig:sample}(b). In addition to the strain beneath four adjacent gates, $\sigma_x$ and $\sigma_z$ are enhanced along vertical stripes with the period of $\lambda/2=251.6\,$nm, which indicate an SSAW. $\sigma_y$ shows no such periodic signal.
(b) Contributions of the SSAW, $\delta\sigma_x^2(x)$ and $\delta\sigma_z^2(x)$, to the variances averaged along $y$ within the regions indicated by horizontal dashed lines in panel (a). The solid lines are the measurements, the dashed lines (blue) are model predictions for $C=25\,$pm. The model line of $\sigma_z$ is reduced by one order of magnitude. A vertical solid line facilitates the comparison of the phases between $\delta\sigma_x^2(x)$ and $\delta\sigma_z^2(x)$.}
\label{fig:sigma}
\end{figure*}

In \fig{fig:sigma}{a}
we present the components $\sigma_i(x,y)$ with $i=x,y,z$ for a high resolution SXDM scan of the center of the cavity. The average values of $\sigma_x$ and $\sigma_y$ are similar while that of $\sigma_z$ is smaller. This leads to a distortion of the unperturbed reflection which can be seen in the two inserts at the top of \fig{fig:sigma}{a}. The strong signals below the Ti/Al-gates indicate distortions of the GaAs lattice, which likely occurred during cool-down after the evaporation of the metals at an elevated temperature. This is an interesting observation in itself but not a topic of this article. In addition, in the central region between the gates (highlighted by a pair of dashed lines) periodic oscillations of $\sigma_{x,z}(x)$ appear. They have a period of $\lambda/2=251.6\,$nm and disappear if no SSAW is generated. As expected for Rayleigh waves no such oscillations occur for $\sigma_{y}$.

For an analysis we model the SSAW near the center of the cavity as a superposition of a Rayleigh wave propagating in $x$-direction and its reflection from a Bragg mirror. We consider the non-vanishing components of a Rayleigh wave, the longitudinal compression wave in $x$-direction and the transversal shear wave in $z$-direction. Assuming fixed ends for both relevant strain directions the SSAW is characterized by the displacement field
\begin{align}\label{eq:displacement}
\mathbf{u}(x,z,t)=-2\begin{pmatrix}\chi(z) \\ \zeta(z)\end{pmatrix} \begin{pmatrix}\cos(\omega t) \\ \sin(\omega t)\end{pmatrix}\sin(kx)\,.
\end{align}
In \app{app:rayleigh} we discuss alternative reflection conditions. The functions $\chi(z)=2C\text{e}^{-\Omega_\text{r} k z}\cos (\Omega_\text{i} k z+\varphi)$ and $\zeta(z)=2C|\gamma|\text{e}^{-\Omega_\text{r} k z}\cos(\Omega_\text{i} k z+\varphi+\theta)$ with material-dependent parameters $\Omega=\Omega_\text{r}+\emph{i}\Omega_\text{i}$, $\gamma=|\gamma|\text{e}^{-\emph{i}\theta}$, and $k=2\pi/\lambda$ describe the penetration of the Rayleigh wave by about one wavelength into GaAs \cite{Schuetz2015}. The constant $C$ is proportional to the amplitude of the SSAW. A straightforward calculation based on infinitesimal strain theory, cf.\ Appendices \ref{app:diffraction} and \ref{app:rayleigh}, yields the time averaged modulation, $\delta\boldsymbol{q}$, of the diffraction vector caused by an individual scattering event as a function of $x$ and $z$
\begin{equation}
\begin{split}\label{eq:sigma}
\left<\delta q_x^2\right>_t&=2q_x^2\left[ \chi(z)  k\, \cos (kx) + \frac{q_x}{q_z}\,\frac{\partial \chi(z)} {\partial z}\sin (kx) \right]\\
\left<\delta q_z^2\right>_t&=2q_z^2\left[ \frac{q_z}{q_x}\,\zeta(z) k\, \cos (kx) + \frac{\partial \zeta(z)}{\partial z}\sin (kx) \right]\,.
\end{split}
\end{equation}

In \app{app:numerical} we have numerically summed up the contributions of all scattered photons to the actual reflection to simulate $\sigma_{x,z}(x)$. Here we present an instructive and simplifying approximation, cf.\ \app{app:approx_model} which predicts identical results as our numerical simulation in the phases of $\sigma_{x,z}$ and the ratio $\sigma_x/\sigma_z$ while the pre-factor $C$ differs by $<10\,\%$. The SSAW causes an additional local broadening of the unperturbed reflection by a tiny amount $\sigma_{x,z}(C)/\sigma_{x,z}(0)-1\sim10^{-5}$. Such a small perturbation motivates us to apply an incoherent model of statistically independent contributions of individual scattering events to the broadening of the reflection. Then, the measured variances can be expressed as $\sigma_i^2=(\sigma_0)_i^2+\delta\sigma_i^2$, where $(\sigma_0)_i$ indicates the standard deviations of the unperturbed beam and $\delta\sigma_i$ the small contributions of the SSAW. Because of $\delta \sigma_i^2\ll\sigma_i^2$ we can further express the sum of the contributions of all scattering events along the path of the incident X-ray beam as $\delta\sigma_i^2=\sum\left<\delta q_i^2\right>_t$.

Integration along the incident beam finally results in the model curves for the SSAW contributions, $\delta\sigma_{x,z}^2$, plotted as dashed lines in \fig{fig:sigma}{b} using the amplitude $C=25\,$pm as the only fit parameter. The vertical line near $x=-1\,\mu$m helps visualizing a phase shift between $\delta\sigma_{x}^2$ and $\delta\sigma_{z}^2$. The corresponding experimental results are presented as solid lines, where we subtracted $(\sigma_0)_i^2$ using $(\sigma_0)_x=29.1\,\mu\text{m}^{-1}$ and $(\sigma_0)_z=5.5\,\mu\text{m}^{-1}$ from the undisturbed reflections. It is remarkable that our model data confirm the measured phase shift. For the chosen value of $C$ the model agrees with the observed $\delta\sigma_x$, while the measured $\delta\sigma_z$ is an order of magnitude smaller compared to the prediction. We attribute this deviation to limitations in the quantitative analysis of the amplitude of the measured $\delta\sigma_z$, related with the finite resolution of the X-ray scan, such that only few points were detected across the narrow width $(\sigma_0)_z$ of the reflection. Note, that the accuracy of the amplitude of $\sigma_x$ as well as the phases of both components are not affected by these limitations, which could be avoided in an optimized future SXDM experiment. Given the absence of fit parameters the agreement between measured and model data regarding phase shift is remarkable good.

\begin{figure}[t]
\includegraphics[width=0.85\columnwidth]{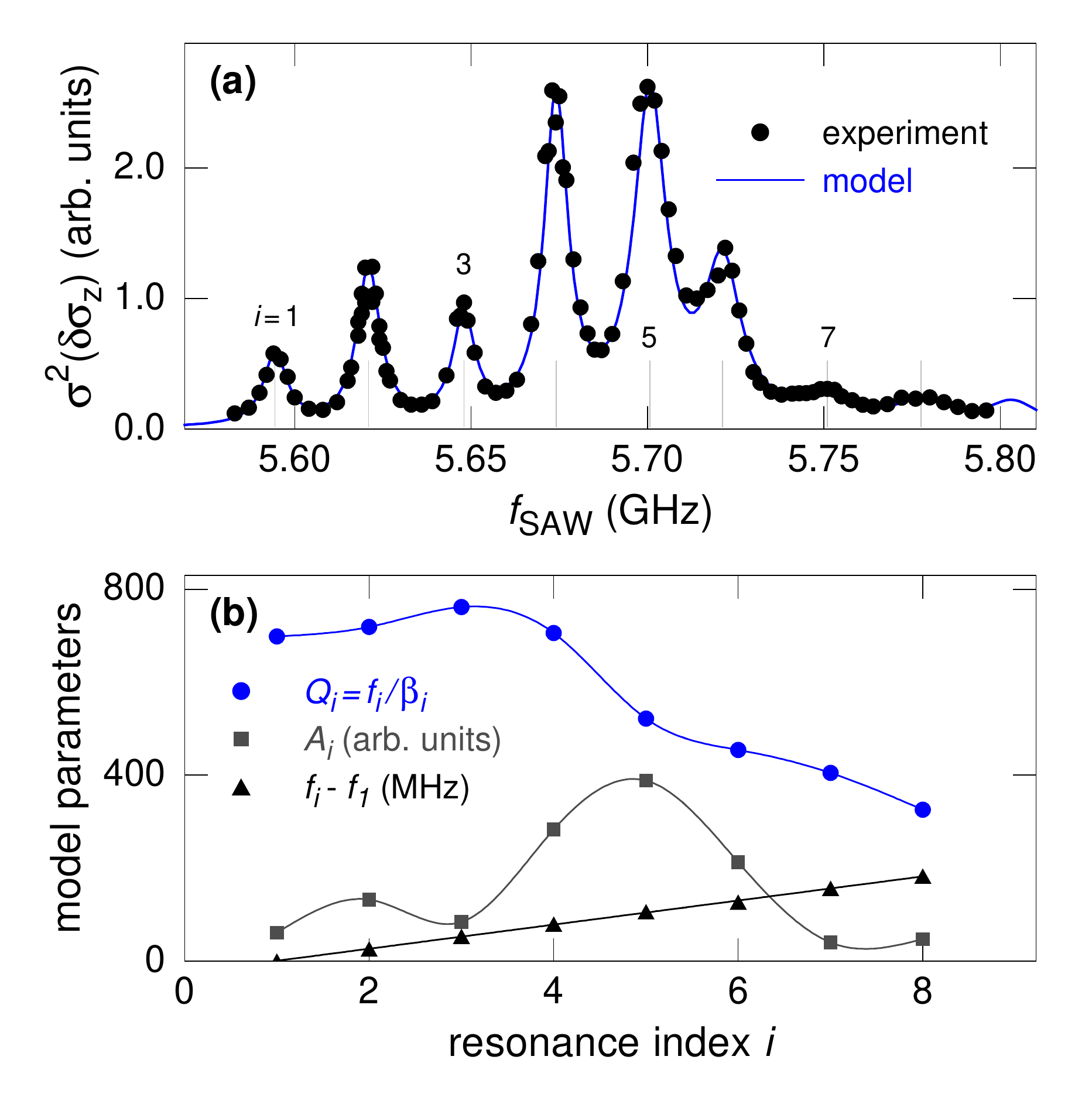}
\caption{(a) Variance of $\delta\sigma_z$ along line scans as shown in Fig.\ \ref{fig:sigma} as a function of frequency; $\sigma(\delta\sigma_z)$ is proportional to the amplitude of the SAW. The solid line is the sum of Lorentzians $\sum_i A_i\beta_i f_\text{SAW}\,/\,((f_\text{SAW}^2-f_i^2)^2+(\beta_if_\text{SAW})^2)$. (b) Fit-parameters used for the model curve in (a): quality factors $Q_i=f_i\,/\,\beta_i$ (circles), amplitudes $A_i$ (squares), and resonance frequencies $f_i-f_1$ (triangles) as a function of the resonance index number which we set to $i=1$ at the lowest frequency measured cavity mode.}
\label{fig:frequency}
\end{figure}
So far we focused on the strain field of the SSAW. However, the SXDM measurements are also suited for characterizing the cavity modes. Larger area surface scans as the one shown in \fig{fig:sample}{b} give an overview of a mode's geometry. In \fig{fig:frequency}{a} we present the frequency dependence, $\sigma_{z}(\fsaw)$, within $5.58\,\text{GHz}\le\fsaw\le5.81\,$GHz determined from high resolution surface scans as in \fig{fig:sigma}{} in the center of the cavity. The almost equidistant maxima correspond to the cavity modes. The mode spacing of $\delta f\simeq25\,$MHz corresponds to an effective cavity length of $L_\text{eff}=c/2\delta f\simeq 56\,\mu$m, which exceeds the distance between the innermost Bragg mirror fingers by $6\,\mu$m (where we used $c=\lambda f\simeq500\,\text{nm}\times5.68\,\text{GHz}\simeq2840\,$m/s).  The strong decrease of the peak amplitude towards the highest frequency of $5.81\,$GHz indicates the high-frequency end of the stop-band of the Bragg mirrors. The solid line is a fit of a sum of Lorentz curves with the fit-parameters plotted in \fig{fig:frequency}{b}. The $Q$-factor (triangles) of the cavity reaches 750, corresponding to a phonon dephasing time of $47\,\mu$s, but decreases towards the edge of the stop-band. The squares indicate the approximate area below each peak, which is proportional to the energy stored in the corresponding mode. It has a maximum around $f=5.68\,$GHz. SXDM provides a direct method for determining the true spectrum of the an rf-SAW cavity with a striking resolution.

\section{Summary and Outlook}\label{sec:summary}

We present the first direct imaging study of the three-dimensional strain field of a rf SSAW at a frequency near 6\,GHz. Even for studies of SAWs of comparable frequency, this goes far beyond what alternative methods can provide, which are at most sensitive to the displacements perpendicular to the surface. Based on our quantitative extraction of strain through spatially resolved Bragg diffraction at the GaAs(335) reflection we are able to show a match of the strain field generated by the SSAW as well as its transversal and longitudinal surface displacements with the theory of Rayleigh waves. Our measurements reveal ---in agreement with the model--- a relative phase shift between the components of the reflection broadening along the propagation direction and perpendicular to the surface. Beyond, we demonstrate the capability of scanning X-ray diffraction microscopy to fully characterize an SSAW and its generating cavity. Our results pave the way towards new applications in future quantum technologies, where the complete knowledge of the three dimensional local strain field of cavity phonons including their phases will be crucial. An example is the coupling between a coherent phonon field and a solid state quantum bit, which sensitively depends on the local amplitude and phase of the strain field components.

%running wave: time average of first momentum disappears at every point, time average of second momentum is constant
%standing wave: time average of first momentum disappears, time average of $\sigma$ oscillates in space

%\subsection{Standing waves in Cavity}
%\begin{itemize}
%\item
%choosing data from two comparable samples with identical dimensions but either with or without Bragg reflector to get a comparison between both. (Maybe it would be better to first focus on the sample without additional Bragg reflectors and present the second sample only where discussing the corresponding data?)

%\item
%proof of principle for resolving standing waves at 251.6\,nm period, fig.\,\ref{fig:sigma}, discussion of different geometries [symmetric (004) vs. asymmetric (335)]

%\item
%phase shift $\sigma_{xx}$ vs. $\sigma_{zz}$. Do FEM calculation, fig.\ref{fig:frequency} confirm this? What does it mean that the standard deviation of the fitted Gaussians are phase-shifted? Due to symmetry reasons $\sigma_{yy}$ stays rather insensitive.
%\end{itemize}

%\subsection{Mode pattern within IDT}
%\begin{itemize}
%\item
%sample 7U with Bragg reflector (small cavity) shows a complex and strongly frequency-dependent strain pattern while crossing the resonance at 5.690\,GHz during a frequency sweep, figs.\,\ref{fig05} and \ref{fig06}.
%\end{itemize}
%\subsection{sample 1L ($\lambda$=500\,nm, without Bragg reflector)}

\section*{Acknowledgement}

The authors thank C.\ David and F.\ Koch from the Paul Scherrer Institut for providing an exceptionally good focusing optics for the X-ray beam, W.\ Anders and A.\ Tahraoui for the fabrication of the SSAW cavities, A. Kuznetsov for support with preliminary S-parameter measurements, S. F\"olsch for an internal review of the manuscript and C.\ B\"auerle for providing us with an rf generator in Grenoble. We highly appreciate beamtime access through project MA-4449 at the ID01 beamline of the European Synchrotron Radiation Facility. This work was financially supported by the Deutsche Forschungsgemeinschaft through grants SA 598/15-1 and LU 819/11-1.

\section*{Contributions of the authors}

M. Hanke, S. Ludwig and P. Santos planned the project. M.E. Msall, P. Santos, J. Hellemann and S. Ludwig designed the rf focusing IDTs and cavities; N. Ashurbekov and S. Ludwig prepared the samples for the SXDM measurements; N. Ashurbekov and P. Santos characterized the electrical rf-properties of the IDTs; E. Zatterin and T.U. Schulli performed the SXDM measurements (locally) in online sessions in collaboration with N. Ashurbekov,  M. Hanke and S. Ludwig (online). E. Zatterin processed and analyzed the raw data. M. Hanke and S. Ludwig developed the model, performed numerical calculations as well as further data analysis and prepared the manuscript.

\section*{References}

\appendix

\section{Modulation of the diffraction vector by a displacement field}\label{app:diffraction}

In the following we derive the modulation $\delta\mathbf{q}(x,z)$ caused by an SSAW of the scattering vector $\mathbf{q}$ of the GaAs(335) reflection used in our SXDM experiment. As in the main article, we define the $x$-axis to be parallel to the propagation direction [110] and the $z$-axis to be directed in [00$\overline1$], which points inside the crystal perpendicular to the (001)-surface. In this notation, the diffraction vector is

\begin{equation}\label{eq:qvector}
\mathbf{q}=
\begin{pmatrix}
 q_x \\ q_z
\end{pmatrix}
=2\pi
\begin{pmatrix}
 {1}/{a_x} \\ {1}/{a_z}
\end{pmatrix}
=\frac{2\pi}{a}
\begin{pmatrix}
 3\sqrt2 \\ -5
\end{pmatrix}
\end{equation}
with the lattice constant of GaAs $a=5.6533\,$\AA, while $q_y=0$. The reciprocal of $\mathbf{q}$, $\mathbf{a}=(a_x,a_z)^\text{T}=a(\sqrt{2}/6,-1/5)^\text{T}$, is normal to the Bragg planes and its absolute value of $\simeq1.7\,$\AA\ is twice the distance between adjacent Bragg planes.

\begin{figure}[ht]
\includegraphics[width=0.6\columnwidth]{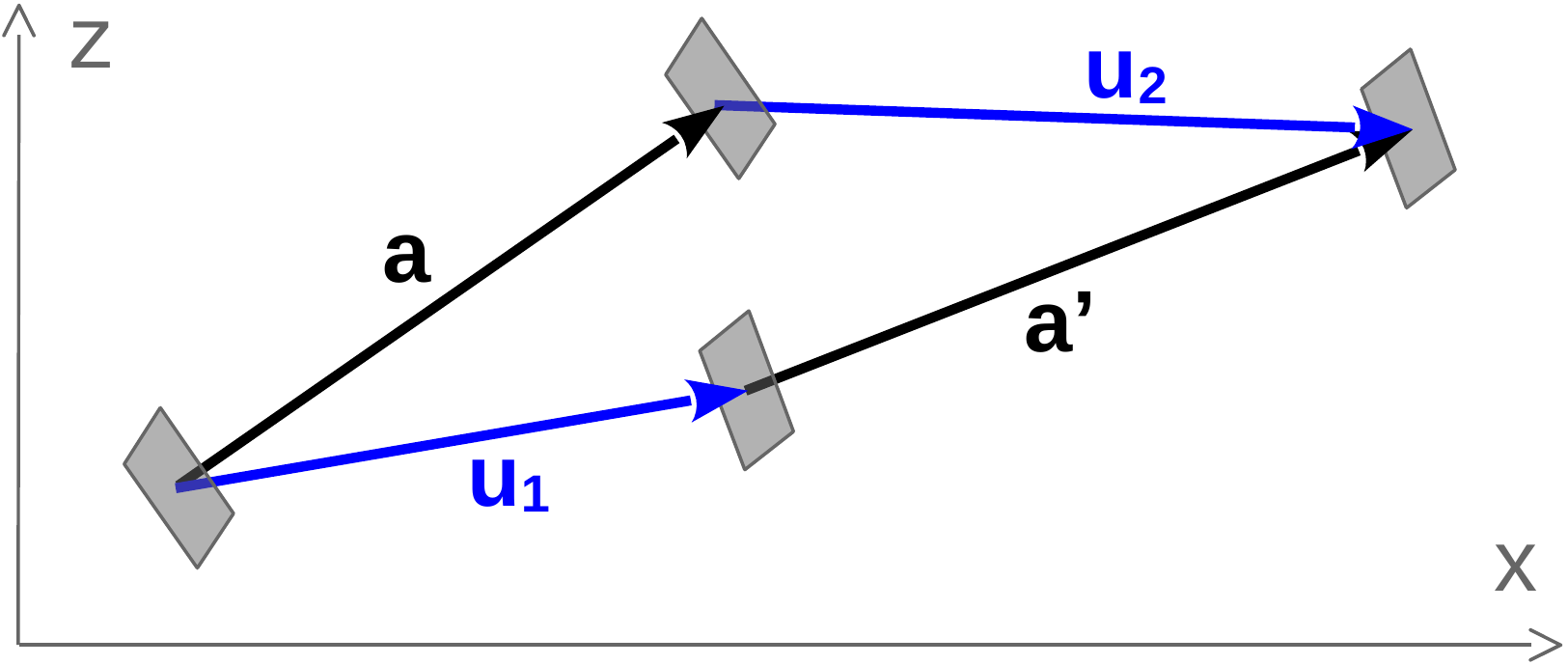}
\caption{Vectors $\mathbf{a}$, $\mathbf{a'}$ connecting adjacent Bragg planes, and displacements $\mathbf{u_1}$ and $\mathbf{u_2}$ of adjacent Bragg planes. The Bragg planes, which without strain are orthogonal to $\mathbf{a}$, are shifted by $\mathbf{u_1}$ and $\mathbf{u_2}$ and tilted, such that they become orthogonal to $\mathbf{a'}$. }
\label{fig:delta_a}
\end{figure}

To determine the modulation of the diffraction vector caused by the displacement field of the SSAW we apply the concept of infinitesimal strain theory. We are interested in the variation $\delta\mathbf{q}=\mathbf{q'}-\mathbf{q}$ corresponding to a variation $\delta\mathbf{a}=\mathbf{a'}-\mathbf{a}$. In \fig{fig:delta_a}{}
we sketch the connection between the displacement vector field $\mathbf{u}$ of the SSAW and the vector $\mathbf{a}$. The displacement field shifts and tilts the Bragg planes such that $\mathbf{a}\rightarrow\mathbf{a'}$. From the sketch we can immediately see that $\delta\mathbf{u}=\mathbf{u_2}-\mathbf{u_1}=\mathbf{a'}-\mathbf{a}=\delta\mathbf{a}$.

To determine the variation $\delta\mathbf{q}$ of the diffraction vector it is convenient to start from its total differential
\begin{equation}
 \text{d}\mathbf{q}=\frac{\partial\mathbf{q}}{\partial x}\text{d}x + \frac{\partial\mathbf{q}}{\partial z}\text{d}z\,.
\end{equation}
Assuming constant strain on the lengthscale of $|\mathbf{a}|$ (infinitesimal strain theory), we can replace $dx\rightarrow a_x$ and $dz\rightarrow a_z$ and write
\begin{equation}
 \delta\mathbf{q}=\frac{\partial\mathbf{q}}{\partial x}a_x + \frac{\partial\mathbf{q}}{\partial z}a_z\,.
\end{equation}
Inserting the components $q_{x,z}$ from \eq{eq:qvector} while using $\partial u_i/\partial j = \partial a_i/\partial j$ (because $\delta\mathbf{u}=\delta\mathbf{a}$) we find
\begin{eqnarray}\label{eq:strainfield}
 \delta q_x &= -q_x\left(u_{xx} + \frac{q_x}{q_z}u_{xz} \right)\nonumber\\
 \delta q_z &= -q_z\left(u_{zz} + \frac{q_z}{q_x}u_{zx} \right)\,,
\end{eqnarray}
where we introduced the notation $u_{ij}=\partial u_i / \partial j$.

\section{Standing Rayleigh waves on a GaAs wafer}\label{app:rayleigh}

SAWs on a homogeneous crystal are Rayleigh waves which can be described as a combination of a compression wave in the propagation direction and a shear wave perpendicular to the surface. The corresponding displacements of a Rayleigh wave can be written as
\begin{equation}\label{eq:rayleigh}
\mathbf{u}
=\begin{pmatrix}u_{x}\\u_{z}\end{pmatrix}
=\begin{pmatrix}\phantom{-}\chi(z) \sin(\omega t-kx)\\-\zeta(z)\cos(\omega t-kx)\end{pmatrix}\,.
\end{equation}
For GaAs the functions $\chi(z)$ and $\zeta(z)$, which describe the penetration of the SAW into the crystal, are \cite{Schuetz2015}
\begin{equation}\label{eq:prefactors}
\begin{split}
\chi &=2C\text{e}^{-\Omegar kz}\,\cos\left(\Omegai kz+\varphi\right)\\
\zeta&=2C|\gamma|\text{e}^{-\Omegar kz}\,\cos\left(\Omegai kz+\varphi+\theta\right)\,
\end{split}
\end{equation}
with $\Omega=\Omegar+i\Omegai\simeq0.5+0.48i$, $\gamma=|\gamma|\text{e}^{-i\theta}\simeq1.345\text{e}^{-4.182i}= -0.68+1.16i$ and $\varphi\simeq1.05$.
Directly at the surface $\chi(z)$ and $\zeta(z)$ are both positive, which results in a retrograde elliptical motion of surface elements for a propagating Rayleigh wave.

A standing wave can be described as the superposition of a left-moving and a right-moving wave of equal amplitudes and frequencies, where their phase differences depend on the reflection conditions at the ends of the cavity. Since the SAW penetrates into the Bragg mirror, it is convenient to introduce an effective cavity length $L_\text{eff}$, which exceeds the distance between the innermost mirror gates $L$. In the main article, we determined $L_\text{eff}=56\,\mu$m from the frequency separation of the modes, which corresponds to a penetration depth of $\delta L\equiv(L_\text{eff}-L)/2\simeq6\lambda$.

Our cavity is mirror symmetric, cf.\ \fig{fig:sample}{a}. As a consequence, the standing wave modes of the cavity alternate in being mirror versus point symmetric (corresponding to even versus odd parity) in respect to the center of the cavity. Further, we can describe the reflection conditions in terms of open or fixed ends. For $\lambda\ll L_\text{eff}$, as is the case for our cavity, the mode number $2L_\text{eff}/\lambda$ is large and, hence, the choice of open versus fixed ends is irrelevant for the observed standing wave modes. However, the phase difference between the two SSAW induced strain components, e.g., $\delta\sigma_x$ versus $\delta\sigma_z$ depends on whether the two components $u_x$ and $u_z$ have the same or opposite parity. (In respect to the center of a mirror symmetric cavity, $u_x$ and $u_z$ have identical parity for equal reflection conditions or opposite parity, if $u_x$ and $u_z$ have opposite reflection conditions.) This leaves us with two choices, which result in qualitatively different standing Rayleigh waves: Either $u_x$ and $u_z$ have equal reflection conditions (both ends fixed or both ends open) or $u_x$ and $u_z$ have opposite reflection conditions ($u_x$ has fixed ends and $u_z$ open ends or vice versa).

\subsection{SSAW with fixed ends for $u_x$ and for $u_z$}\label{app:fixed-fixed}

In the main article, we considered the case of equal reflection conditions, say fixed ends for both SAW components $u_{x}$ and $u_{z}$. The corresponding standing wave has the form
\begin{align}\label{eq:SSAW}
\mathbf{u}=-2\begin{pmatrix}\chi(z) \\ \zeta(z)\end{pmatrix} \begin{pmatrix}\cos(\omega t) \\ \sin(\omega t)\end{pmatrix}\sin(kx)\,.
\end{align}
The strain field of the SSAW for fixed ends only described by the partial derivatives of the displacements is
\begin{eqnarray}\label{eq:derivatives}
  u_{xx}&=-2\chi &k \cos(\omega t)\cos(kx)\nonumber\\
  u_{xz}&=-2\frac{\partial\chi}{\partial z} & \cos(\omega t)\sin(kx)\nonumber\\
  u_{zx}&=-2\zeta &k \sin(\omega t)\cos(kx)\nonumber\\
  u_{zz}&=-2\frac{\partial\zeta}{\partial z} & \sin(\omega t)\sin(kx).
\end{eqnarray}
Inserting \eq{eq:derivatives} into \eq{eq:strainfield}, we find the local variations of the diffraction vector caused by the SSAW
\begin{equation}
\begin{split}\label{eq:sigma_appendix}
\delta q_x&=2q_x\left[                \chi(z)k \, \cos (kx) + \frac{q_x}{q_z}\frac{\partial\chi(z)} {\partial z}\sin (kx) \right] \cos(\omega t)\\
\delta q_z&=2q_z\left[\frac{q_z}{q_x} \zeta(z)k\, \cos (kx) +                \frac{\partial\zeta(z)}{\partial z}\sin (kx) \right] \sin(\omega t)\,.
\end{split}
\end{equation}
Taking the time average of the square of \eq{eq:sigma_appendix} yields \eq{eq:sigma} in the main article.
By inserting \eq{eq:prefactors} these variations can be written in the explicit form
\begin{widetext}
\begin{equation}\label{eq:sigma_final_equal}
\begin{split}
\delta q_x&=2q_x k \chi(z)  \left[ \phantom{\frac{q_z}{q_x}} \cos (kx) - \,\frac{q_x}{q_z} \,\Big(\Omegar+\Omegai\tan(\Omegai k z+\phi) \Big)\;\sin (kx)\right] \cos(\omega t)\\
\delta q_z&=2q_z k \zeta(z) \left[          \frac{q_z}{q_x}  \cos (kx) - \Big(\Omegar+\Omegai\tan(\Omegai k z+\phi+\theta) \Big)\sin (kx)\right] \sin(\omega t)\,.
\end{split}
\end{equation}
\end{widetext}

\subsection{SSAW with fixed ends for $u_x$ and open ends for $u_z$}\label{app:fixed-open}

Next we perform the same calculation as in \sect{app:fixed-fixed} but assuming a standing Rayleigh wave with fixed ends for $u_x$ and open ends for $u_z$. It has the form
%^
\begin{align}\label{eq:SSAW-2}
\mathbf{u}(\mathbf{r},t)=-2\begin{pmatrix}\chi(z) \\ \zeta(z)\end{pmatrix} \begin{pmatrix}\sin(kx) \\ \cos(kx)\end{pmatrix}\cos(\omega t)\,.
\end{align}

The strain field of the SSAW is described by the partial derivatives of the displacements
\begin{eqnarray}\label{eq:derivatives-2}
  u_{xx}&=-2\chi &k \cos(\omega t)\cos(kx)\nonumber\\
  u_{xz}&=-2\frac{\partial\chi}{\partial z} & \cos(\omega t)\sin(kx)\nonumber\\
  u_{zx}&=\phantom{-}2\zeta &k \cos(\omega t)\sin(kx)\nonumber\\
  u_{zz}&=-2\frac{\partial\zeta}{\partial z} & \cos(\omega t)\cos(kx).
\end{eqnarray}
Inserting \eq{eq:derivatives-2} into \eq{eq:strainfield} we find
\begin{equation}
\begin{split}\label{eq:sigma_appendix-1}
\delta q_x&=2q_x\left[                \chi(z)k \, \cos (kx) + \frac{q_x}{q_z}\frac{\partial\chi(z)} {\partial z}\sin (kx) \right] \cos(\omega t)\\
\delta q_z&=2q_z\left[\frac{\partial\zeta(z)}{\partial z}\cos (kx) -                \frac{q_z}{q_x} \zeta(z)k\, \sin (kx) \right] \cos(\omega t)\,.
\end{split}
\end{equation}
The component $\delta q_x$ is identical as for two fixed ends, while $\delta q_z$ is shifted in phase, both along the $x$-axis and in time.
By inserting \eq{eq:prefactors} these variations can be written in the explicit form
\begin{widetext}
\begin{equation}
\begin{split}\label{eq:sigma_final_different}
\delta q_x&=\phantom{-}2q_x k \chi(z)   \left[\quad\,         \cos (kx) - \;\frac{q_x}{q_z} \;\Big(\Omegar+\Omegai\tan(\Omegai k z+\phi) \Big)\;\sin (kx)\right] \cos(\omega t)\\
\delta q_z&=-2q_z k \zeta(z)\,\left[\frac{q_z}{q_x} \sin (kx) + \Big(\Omegar+\Omegai\tan(\Omegai k z+\phi+\theta) \Big)\cos (kx) \right] \cos(\omega t)\,.
\end{split}
\end{equation}
\end{widetext}

\section{Model used in the main article}\label{app:approx_model}

In our SXDM measurements we probe the standard deviations $\sigma_{x,z}$ of the broadened reflection $\mathbf{q}$ in reciprocal space. Following the model outlined in the main article, the contribution of the strain field of the SSAW can be expressed as $\sigma_{x,z}^2=\left<\delta q_{x,z}^2\right>_t$, which can be easily calculated analytically starting from \eq{eq:sigma_final_equal} for equal reflection conditions or from \eq{eq:sigma_final_different} for opposite reflection conditions of $u_x$ versus $u_z$. Our last step is adding up all contributions to the actually measured reflection by numerically integrating along the incident beam of photons.

In GaAs, the absorption length of X-rays at the energy of 10.185\,keV is $l_\text{abs}\simeq55\,\mu$m. In comparison, the extinction depth of the X-ray beam for the (335)-reflection is much smaller because of the Bragg reflection, $l_\text{ext}\simeq6.6\,\mu$m. As sketched in \fig{fig:setup}{b} of the main article, a photon of the X-ray beam penetrates into the crystal until it is randomly scattered at a penetration length $l_\text{i}=z/\sin\alpha_\text{i}\simeq1.004z$. Then the photon travels for the distance $l_\text{o}=z/\sin\alpha_\text{f}\simeq12.5z$ before it leaves the crystal. For such a diffraction process at a given depth $z(l_\text{i})$ below the surface the effects of extinction and absorption can be combined in a damping factor
\begin{equation}\label{eq:damping}
\beta(z)=\text{e}^{-[l_\text{i}/l_\text{ext}+(l_\text{i}+l_\text{o})/l_\text{abs}]}\,.
\end{equation}
For our $\lambda\simeq500\,$nm we find $\beta(z)\simeq\text{e}^{-0.2 z/\lambda}$. The contribution of a diffraction at $z=500\,$nm to the reflection is therefore reduced by approximately 18\%.
The SSAW itself decays much faster, namely within one wavelength below the surface, according to $\chi(\lambda)/\chi(0)\simeq-0.05$ and $\zeta(\lambda)/\zeta(0)\simeq0.1$. To simulate our experiment we integrate $\left<\delta q_{x,z}^2(x_z,z)\right>_t$ along the incident beam trajectory. This integration is particularly relevant because of the oscillating behavior of $\chi(z)$ and $\zeta(z)$ for $z\le\lambda$. Using $x_z=x(z=0)+z/\tan\alpha_\text{i}\simeq x(z=0)+0.084z$ and including the depth dependent attenuation we arrive at our prediction for the contribution of the SSAW to the shape of the reflection
\begin{equation}\label{eq:final}
{\delta\sigma_{x,z}^2}=\frac{\int_0^\infty\text{d}z\left<\delta q_{x,z}^2(x_z,z)\right>_t\beta(z)}{\int_0^\infty\text{d}z\beta(z)}\,.
\end{equation}
%
%
%\begin{equation}\label{eq:final}
%{\delta\sigma_{x,z}^2}\simeq\frac{0.2}{\lambda}\int_0^\infty\text{d}z\left<\delta q_{x,z}^2(x_z,z)\right>_t\text{e}^{-0.2z/\lambda}\,,
%\end{equation}
%
%where the pre-factor $0.2/\lambda$ normalizes the damping factor.
To determine the model curves in \fig{fig:sigma}{d} we computed this integral numerically inserting $\delta q_{x,z}$ from \eq{eq:sigma_final_equal}, which applies for equal reflection conditions of both components of the SSAW.

By applying  \eq{eq:final} we incoherently add up the contributions to the Bragg reflections of infinitesimal volume elements along the incident X-ray beam. This incoherent approach might be questioned, as the penetration depth of the SSAW and the nominal longitudinal X-ray coherence length of about 800\,nm \cite{LCD19} are similar. However, the photon coherence is much more reduced by the lateral focusing of the X-ray beam, which by orders of magnitude dominates the broadening of the reflections and, thus, justifies our incoherent model.

%By applying  \eq{eq:final} we incoherently add up the contributions to the Bragg reflections of infinitesimal volume elements along the incident X-ray beam. This incoherent approach is justified by the following consideration: On the one hand, the penetration depth of the SSAW and the nominal longitudinal X-ray coherence length of about 800\,nm \cite{LCD19} are similar. On the other hand, the spectral width of the focused X-ray beam related to the spectral acceptance window of the monochromator causes additional dephasing, which reduces the effective longitudinal coherence length to a value far below 800\,nm.

\begin{figure}[t]
\hspace{-5mm}
\includegraphics[width=.5\textwidth]{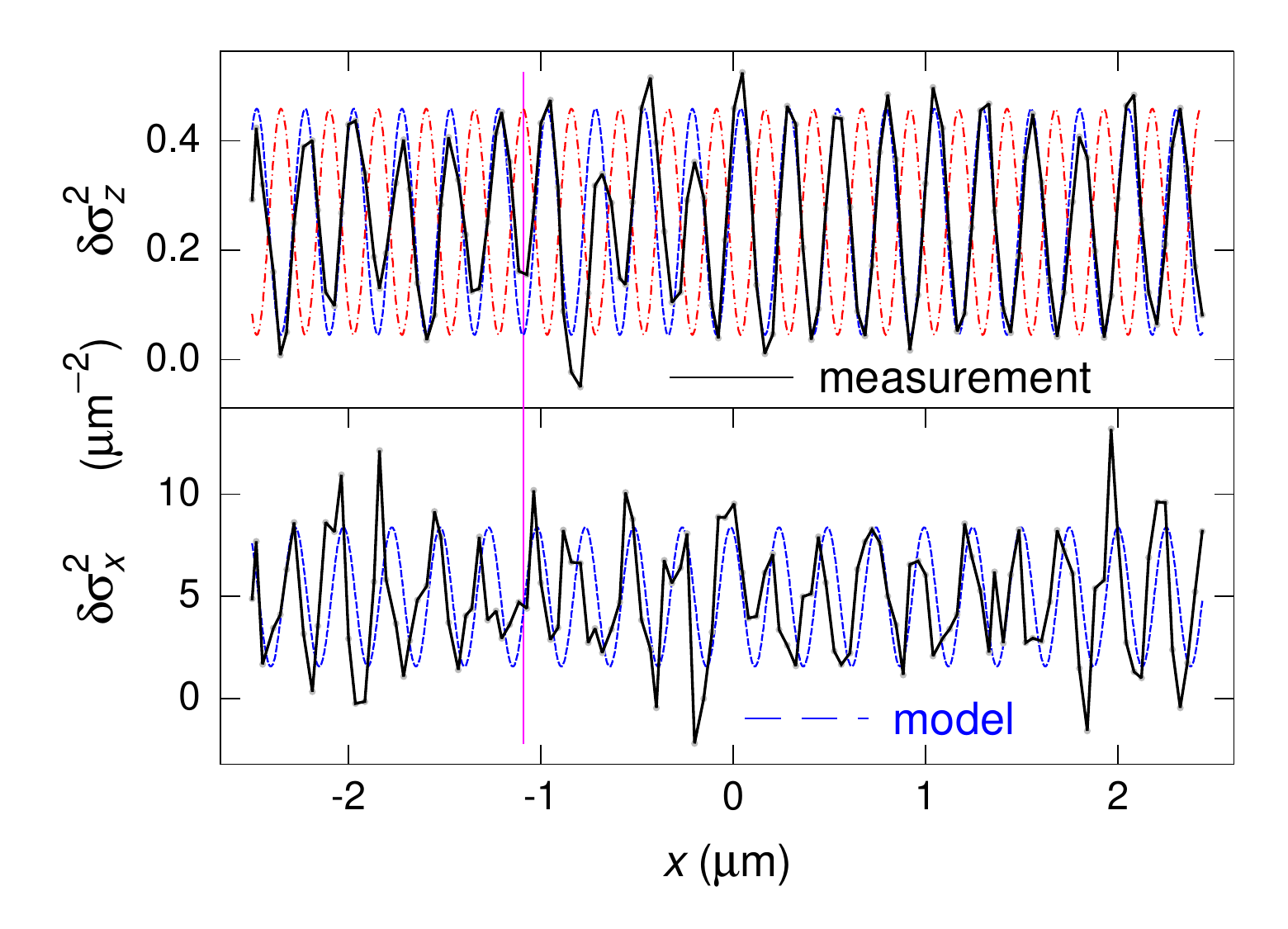}
\caption{Identical measured data as in Fig.\ \ref{fig:sigma}(d) of the main article (solid lines, black). Dashed lines (blue) are model curves for fixed ends for both SSAW components, as in Fig.\ \ref{fig:sigma}(d) of the main article. The dot-dashed line (red) is the corresponding model curve but using open ends for the $z$-component of the SSAW. Its phase is shifted by $\pi$ compared to the case of fixed ends. Model lines of $\sigma_z$ are reduced by a factor 10.}
\label{fig:sigma_comp}
\end{figure}

In \fig{fig:sigma_comp}{}
we compare the two cases of equal versus different reflection conditions of the $x$- and $z$-components of the SSAW, by insertion of \eq{eq:sigma_final_equal} versus \eq{eq:sigma_final_different} into \eq{eq:final}. Because we varied the reflection condition for the $z$-component of the SAW while keeping it unchanged for the $x$-component, we find two model curves for $\delta\sigma_z^2$ with opposite phase but only one model curve for $\delta\sigma_x^2$. Note, that the phase relation between the components is fixed while we have chosen the overall phase for best agreement with the measured data (solid lines). Our measured data are clearly better described by the combination of equal reflection conditions, e.g., both ends fixed.

\section{Numerical simulation of the SXDM experiment}\label{app:numerical}

We tested our model described above by comparing its results to numerical simulations of the SXDM experiment. Experimentally we find, that the projections of the unperturbed three-dimensional reflection to the $q_x$-, $q_y$- and $q_z$-axes of reciprocal space are well described by Gaussian distributions
$\propto\text{e}^{-\left({q_i}/{2(\sigma_0)_i}\right)^2}$.
The displacement field of the SSAW causes variations of $\mathbf{q}(x,z)$. Hence, the reflections of individual scattering events correspond to Gaussian reflections of the same shape but slightly shifted according to
\begin{equation}\label{eq:amplitude}
A_i(x,z,t)\propto\beta(z)\text{e}^{-\left(\frac{q_i-\delta q_i(x,z,t)}{2(\sigma_0)_i}\right)^2}\,,
\end{equation}
where $\delta q_i(x,z,t)$ has to be inserted from \eq{eq:sigma_final_equal} or \eq{eq:sigma_final_different}, depending on the reflection conditions and $\beta(z)$ is the damping factor introduced in \eq{eq:damping} above.

Following the actual experiment, we next determine the shape of the actual reflection by numerically integrating over $z$ (along the incident beam) and then over time
\begin{equation}\label{eq:integral}
\alpha_i(z)\propto\int_{t=0}^{2\pi/\omega}\text{d}t\int_{z=0}^{\infty}\text{d}z\,A_i(x,z,t)\,.
\end{equation}
Finally, we numerically determine the variances $\sigma_i^2$ and, for a direct comparison with our approximate model above, $\delta\sigma_i^2=\sigma_i^2-(\sigma_0)_i^2$. The numerical results resemble our approximate model results one-to-one in the phase difference between $\sigma_x^2$ and $\sigma_z^2$ and the ratio $\sigma_x/\sigma_z$, while the prefactor $C$ differs by $<10\,\%$. This good agreement between the direct numerical calculation and our analytical approximation is expected, because of the small modifications of the undisturbed reflection by the displacement field of the SSAW.

\section{Displacement amplitudes of the SSAW --- classical regime}\label{app:displacement}

The displacement amplitudes of the SSAW at the surface are $U_x=2C\chi(0)\simeq0.8C$ and $U_z=2C|\gamma|\zeta(0)\simeq1.75C$, cf.\ \eq{eq:displacement}, where we used $\chi(0)\simeq0.4$, $\zeta(0)\simeq0.65$ and $|\gamma|\simeq1.345$. Fitting our model described above to our measurements presented in \fig{fig:sigma}{} we find $C\simeq25\,$pm, corresponding to displacement amplitudes at the surface of $U_x\simeq20\,$pm and $U_z\simeq44\,$pm.

From our measurements presented in \fig{fig:sample}{b} and \fig{fig:sigma}{} we can estimate the mode area as $A\simeq56\,\mu\text{m}^2$. This allows us to further estimate the displacement that a single phonon captured inside the cavity would cause, $C_0=\sqrt{\hbar f\lambda/HA}$, where the material constant $H=28.2\times10^{10}\,$N/m$^2$ for GaAs \cite{Schuetz2015}. Using $\lambda=503.2\,$nm and $f\simeq5.675\,$GHz we find $C_0\simeq0.138\,$fm. With $C=\sqrt{N_\text{ph}}C_0$ we can finally estimate the number of phonons contributing to our SSAW, which is $N_\text{ph}\sim10^{10}$. This large number of phonons clarifies that our measurements are probing the classical regime of SSAWs.

The resolution limit of our method makes such a large amplitude necessary. Another limiting factor is the small energy of our resonant phonons, which is about three orders of magnitude smaller compared to that of the dominant room temperature thermal phonons. Nevertheless, as long as the measured SSAW is not distorted by non-linear effects, we expect that the SSAW of a single phonon corresponds to our measurements but scaled down to the amplitude $C_0$.

\end{document}